# ZTF J191814.28+010850.2: An eclipsing cataclysmic-variable candidate in the SW Sextantis period range

Klaus Bernhard[1], Wolfgang Moschner[2], Christopher Lloyd[3], Peter Frank[4]

1) Bundesdeutsche Arbeitsgemeinschaft für Veränderliche Sterne e.V., Germany
American Association of Variable Star Observers (AAVSO), USA
email: klaus.bernhard@liwest.at, ORCID 0000-0002-0568-0020
2) Bundesdeutsche Arbeitsgemeinschaft für Veränderliche Sterne e.V., Germany
email: wolfgang.moschner@gmx.de
3) Department of Physics and Astronomy, University of Sussex, Brighton BN19QH, UK
email: C.Lloyd@sussex.ac.uk, ORCID 0000-0003-4323-1960
4) Bundesdeutsche Arbeitsgemeinschaft für Veränderliche Sterne e.V., Germany
email: frank.velden@t-online.de

**Abstract:** *We present a photometric study of ZTF J191814.28+010850.2, a short-period variable previously classified as an EA-type eclipsing binary. The analysis is based on ZTF observations in the g and r bands and additional time-series CCD photometry obtained on two nights. We derive an eclipse period of 0.1408529(3) d. The phased ZTF light curves show a deep primary eclipse, with an eclipse depth of about 2.8 mag in g and 2.0 mag in r, as well as indications of cycle-to-cycle variations outside eclipse. After excluding the primary eclipse and accounting for the remaining orbital modulation, robust locally weighted fits reveal a broad non-linear long-term brightness variation in both bands, with peak-to-peak ranges of about 0.56 mag in g and 0.46 mag in r. The deep eclipse, short orbital period and irregular variability suggest that the object is a cataclysmic-variable candidate and possibly a member of the SW Sextantis group. Time-resolved spectroscopy is required to confirm or refute the proposed classification and to test possible membership in the SW Sex class.*

## 1 Introduction

Cataclysmic variables (CVs) are interacting binary systems in which a white dwarf accretes matter from a Roche-lobe-filling companion, usually a late-type main-sequence star. In most systems, the transferred material forms an accretion disc around the white dwarf. If the white dwarf has a sufficiently strong magnetic field, however, the accretion flow may instead be channelled directly towards its magnetic poles. Cataclysmic variables show brightness variations over a wide range of timescales, including outbursts, rapid flickering and periodic modulations related to orbital motion or magnetic effects (e.g. [1], [2]).

SW Sextantis stars, commonly abbreviated as SW Sex stars, form a subgroup of novalike cataclysmic variables that was recognized as a distinct class in the early 1990s [3]. In catalogues such as the General Catalogue of Variable Stars [4] and the AAVSO International Variable Star Index [5], they are generally listed as novalike variables rather than as a separate official class. Most SW Sex systems have orbital periods between approximately 0.10 and 0.17 d and contain a bright accretion disc. Their photometric behaviour often includes eclipses, strong flickering and quasi-periodic variations. Their defining properties, however, are mainly spectroscopic and include strong single-peaked emission lines, high-excitation features and absorption components that vary with orbital phase [3]. SW Sex stars appear to be common and may represent the dominant population of cataclysmic variables with orbital periods between about 0.12 and 0.16 d [6].

Another group of novalike variables is formed by the VY Sculptoris stars. These systems usually remain in a bright state but occasionally undergo pronounced fading events caused by a temporary reduction in the mass-transfer rate. Their orbital periods are often similar to those of SW Sex stars, and the two groups are not necessarily mutually exclusive. Several VY Scl systems also show characteristics associated with the SW Sex phenomenon. We recently identified the VY Scl star ZTF J172132.75+445851.0 using an approach similar to that applied to the variable investigated in the present paper [7].

## 2 Observations and data analysis

We identified ZTF J191814.28+010850.2 as a candidate during a systematic visual inspection of objects with very short periods in the ZTF Catalog of Periodic Variable Stars. In this catalog, the object is classified as an EA-type eclipsing variable with a notably short period of 0.1408551 d [8].

The catalogue contains approximately 780,000 classified periodic variables based on observations obtained by the Zwicky Transient Facility, a time-domain survey operating at Palomar Observatory since 2017. ZTF uses a wide-field camera mounted on the 48-inch Samuel Oschin Schmidt Telescope and observes mainly in the g and r passbands, with some additional observations in i. With a limiting magnitude of approximately 20.5 mag, the survey provides a valuable data set for investigations of variable stars [9–11]. The basic properties of ZTF J191814.28+010850.2 are summarized in Table 1.

Table 1: Basic data for ZTF J191814.28+010850.2

| | | |
|---|---|---|
| Identifiers | USNO-A2.0 0900-14543495 | |
| | XPM 182-0628200 | |
| | TIC 177151792 | |
| | 2MASS 19181426+0108504 | |
| | Gaia DR3 4264672417768180224 | |
| | ATO J289.5595+01.1472 | |
| Magnitude Range (r) | 15.3 – 17.7 | |
| Gaia DR3 [12]: | Right Ascension (J2000) | 19 18 14.271* |
| | Declination (J2000) | +01 08 50.3423* |
| | Gaia G magnitude | 15.42 mag |
| | Parallax | 1.9310±0.0358 mas |

* All coordinates are taken from the Gaia DR3 catalogue (http://vizier.u-strasbg.fr/viz-bin/VizieR?-source=I/355). The coordinates (epoch J2000) are computed by VizieR and are not part of the original data from Gaia (note that the coordinates are computed from the positions and the proper motions).

Additional time-series CCD photometry was obtained by Wolfgang Moschner and Peter Frank on 26 August and 15 September 2025 using the remotely operated Chilescope Telescope 1 in Chile [13]. The instrument is a 1.0-m ASA Ritchey-Chrétien telescope with a focal ratio of f/6.8, equipped with an FLI ProLine 16803 CCD camera. The observations were obtained through a Sloan-r filter.

MuniWin [14] and StarCurve (Lienhard Pagel) [15] were used for differential photometry of the frames. Five comparison stars were used to form an ensemble reference by averaging their measured brightnesses (Gaia DR3 4264674689812834432, Gaia DR3 4264673384170641408, Gaia DR3 4264672044140798848, Gaia DR3 4264674930331060736 and Gaia DR3 4264672696975367168).

## 3 Results

The long-term behaviour was characterized separately in the two ZTF passbands. Orbital phases were calculated from the ephemeris given below, and measurements within a circular phase distance of 0.10 from primary eclipse (i.e. phases 0.90–1.00 and 0.00–0.10) were excluded, leaving 477 observations in g and 877 in r. For each band, a four-harmonic Fourier representation of the remaining orbital modulation was fitted and subtracted. The phase-corrected measurements were then combined into nightly medians. A robust locally weighted regression (LOWESS) [16] was applied with a fraction of 0.24, corresponding to approximately 109 observing nights per local fit in the g band and 123 in the r band. This fraction was adopted as a compromise between reducing short-term scatter and retaining the broad structure of the long-term variation. As a sensitivity test, the analysis was repeated using fractions of 0.20 and 0.30. Over the tested range from 0.20 to 0.30, the overall shape of the trends and the location of the broad minimum near HJD 2460150 remained qualitatively unchanged. The resulting peak-to-peak ranges varied only from approximately 0.53 to 0.56 mag in g and from 0.45 to 0.47 mag in r. For the adopted fraction of 0.24, the peak-to-peak ranges are approximately 0.56 mag in g and 0.46 mag in r. The curves show a broad fading in both bands towards minimum light, followed by a partial recovery. These values provide an empirical description of the long-term variation rather than a physical model. Figure 1 shows the phase-corrected measurements and the band-specific LOWESS trends.

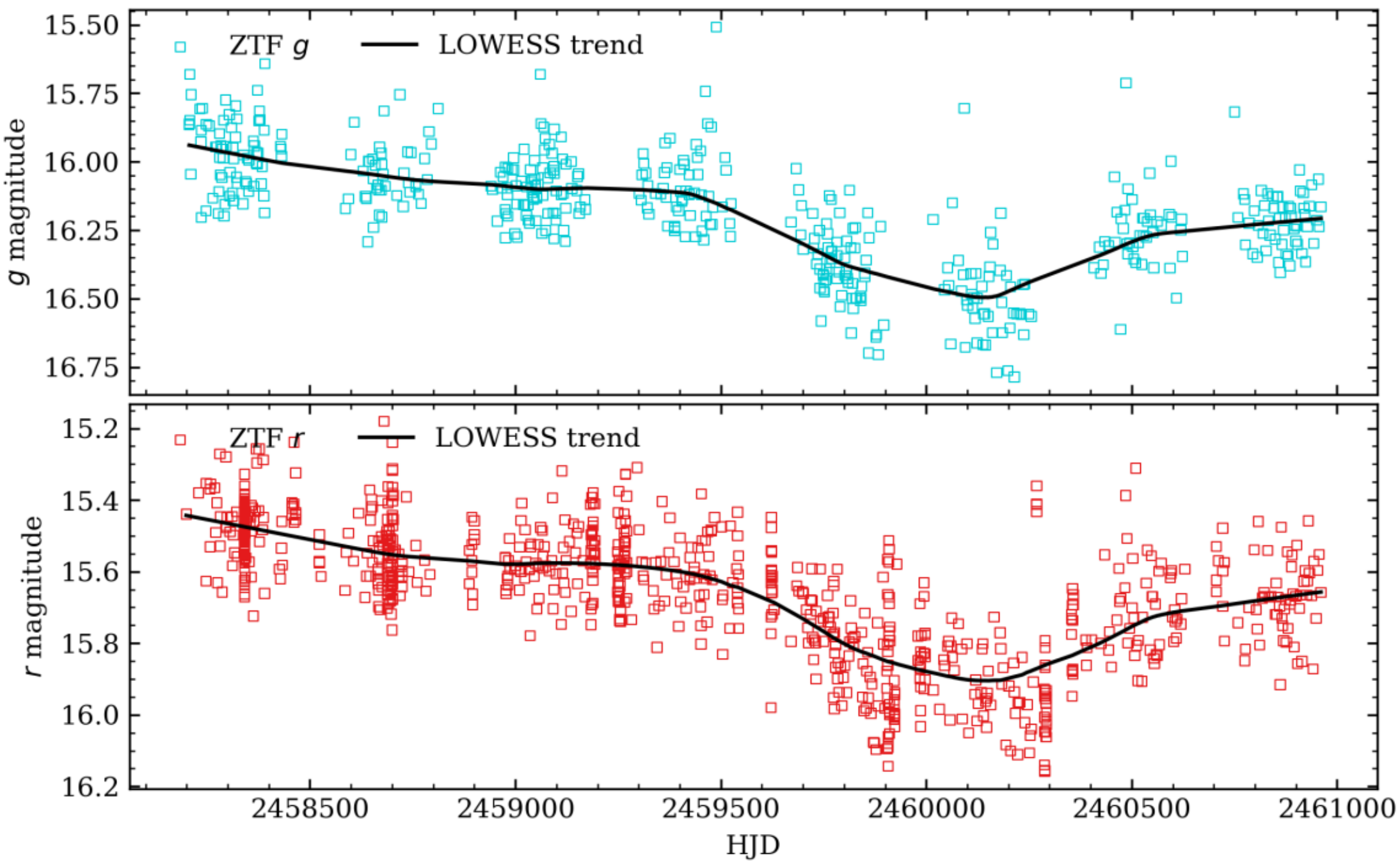

Figure 1: Long-term ZTF light curves in the g and r bands after exclusion of the primary eclipse and subtraction of the fitted orbital modulation. The black curves show robust LOWESS estimates of the non-linear long-term brightness variation.

The period analysis was performed using the ANOVA method implemented in Peranso [17]. The ZTF g-band data set contained 585 usable measurements and was analysed over a period range of 0.01-1.0 d. The strongest signal was found at P = 0.1408529(3) d. No competing alias of comparable significance was found within the searched period range. The ZTF r-band data were folded with the same period and used as a consistency check of the eclipse morphology. The quoted uncertainty represents the formal period uncertainty from the Peranso analysis. As an additional stability check, a combined Fourier analysis gives P = 0.140852974 d for the complete ZTF data set, while separate fits to the early and late halves give 0.140852893 d and 0.140853025 d, respectively. The close agreement of the periods derived from the full data set and the two temporal subsets provides no evidence for a detectable period change within the precision of the present analysis. The resulting ephemeris is:

Min(I) HJD = 2458238.8499(1) + 0.1408529(3) × E

The phased light curves are shown in Figures 2 and 3. Before folding the data, the fitted band-specific LOWESS trends were removed by referring all measurements to HJD 2459000. The light curves display a deep primary eclipse with an eclipse depth of approximately 2.8 mag in the g band and 2.0 mag in the r band. The eclipse profile shows indications of an asymmetry, particularly in the g band, which may indicate asymmetric light contributions from the accretion structure, although its physical origin remains uncertain. The out-of-eclipse brightness is slightly higher between phases 0.0 and 0.5 than during the second half of the orbital cycle, and this structure is persistent through the ZTF data and is typical of novalike systems (see e.g., [6,18,19]). Substantial scatter remains in both detrended light curves. Outside the primary eclipse, the robust residual scatter around the fitted phased light curves is approximately 0.11 mag in g and 0.10 mag in r, several times larger than the median quoted photometric uncertainties of 0.023 and 0.018 mag, respectively. The formal measurement uncertainties alone therefore cannot account for the remaining scatter, which may partly reflect intrinsic cycle-to-cycle variability.

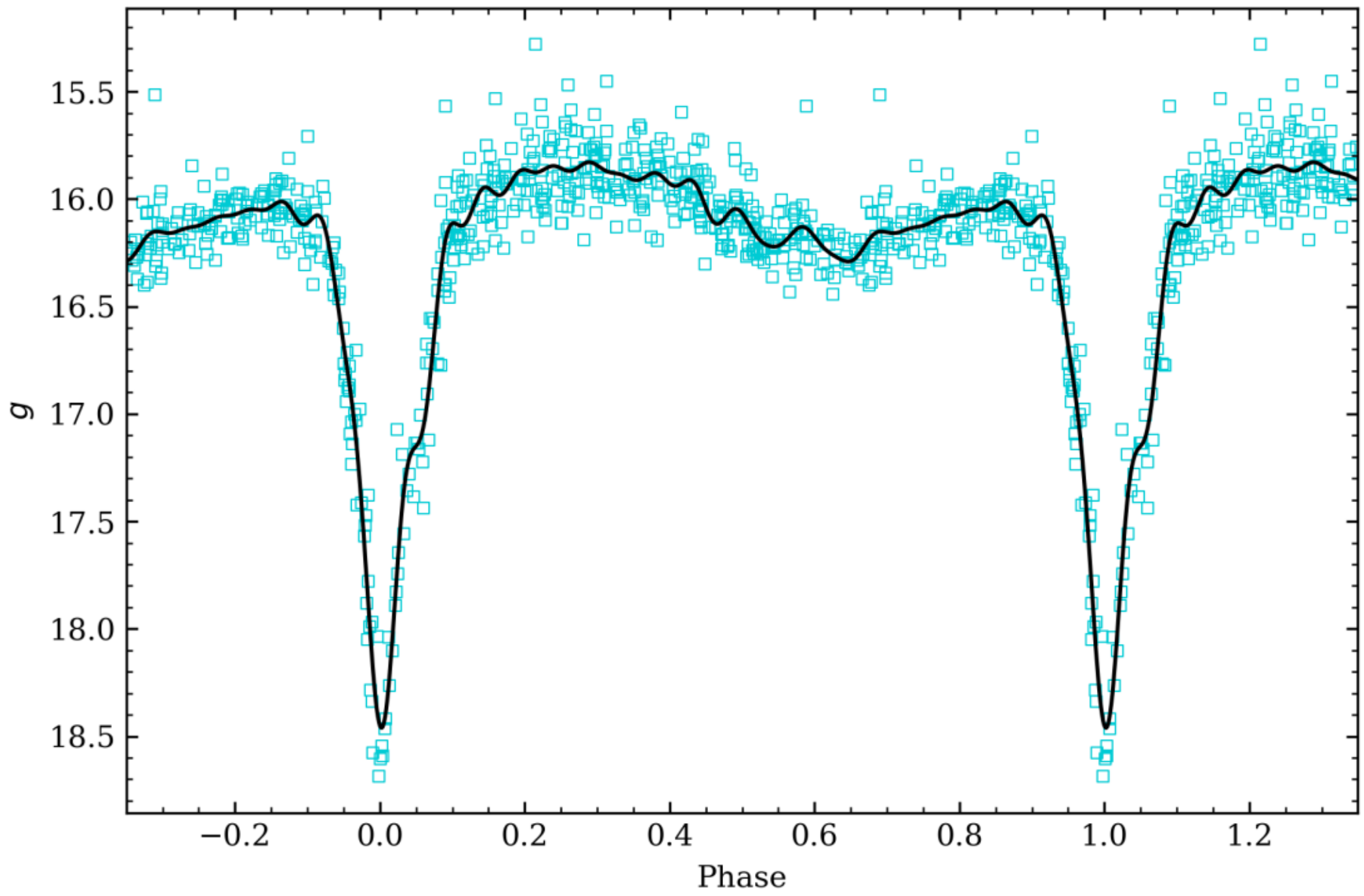

Figure 2: Phased ZTF light curve of ZTF J191814.28+010850.2 in the g band after removal of the smooth long-term variation. The line is a high-order Fourier fit intended to highlight the light-curve morphology and is not a physical model.

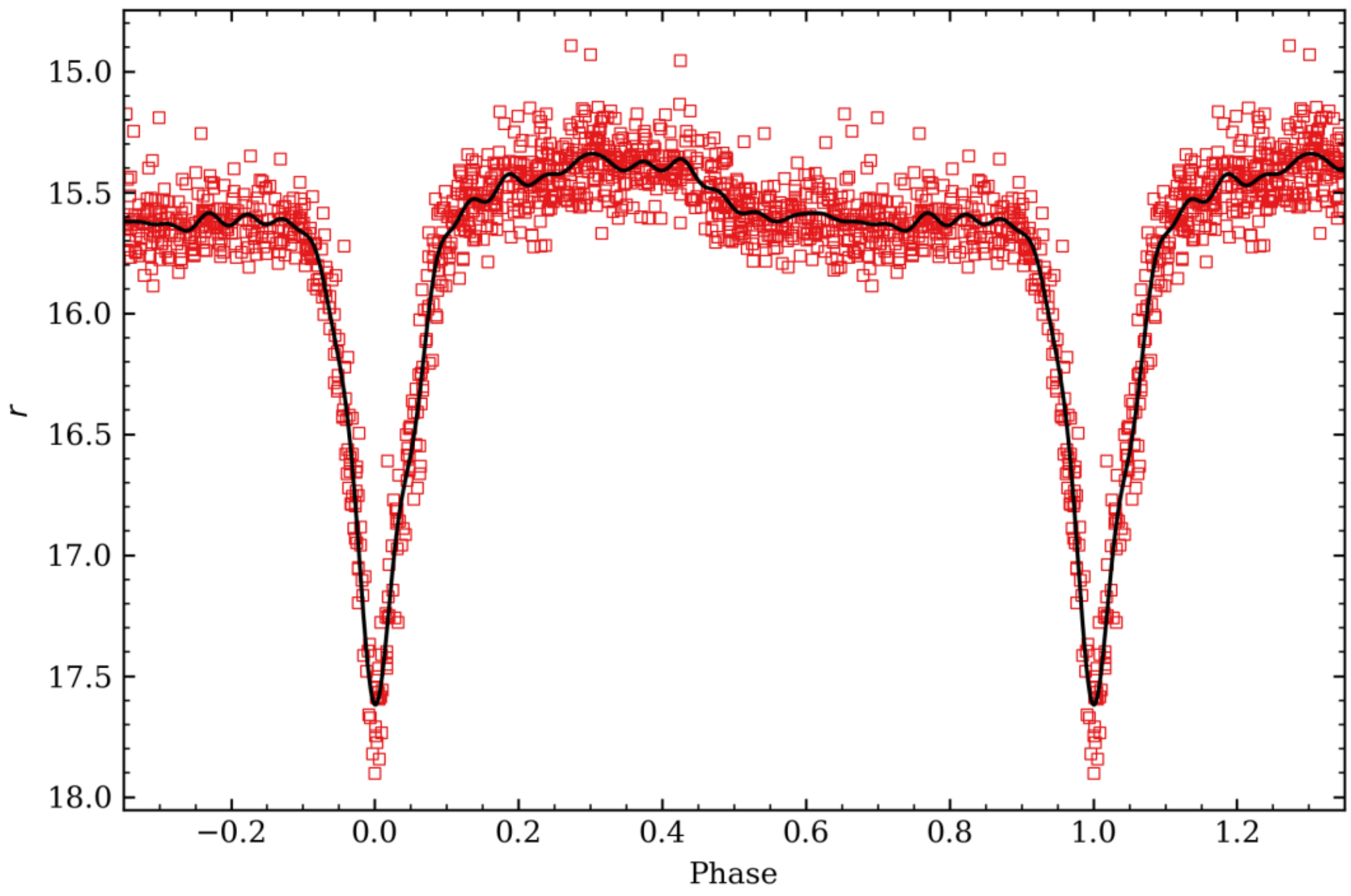

Figure 3: Phased ZTF light curve of ZTF J191814.28+010850.2 in the r band after removal of the smooth long-term variation. The line is a high-order Fourier fit intended to highlight the light-curve morphology and is not a physical model.

The phased light curves from the two observing nights, folded using the calculated ephemeris, are shown in Figure 4. Both data sets clearly confirm the deep primary eclipse. The differences around phase 0.3 may indicate cycle-to-cycle variations, but additional time-series photometry is required to establish whether these differences are intrinsic rather than caused by photometric or reduction-related effects.

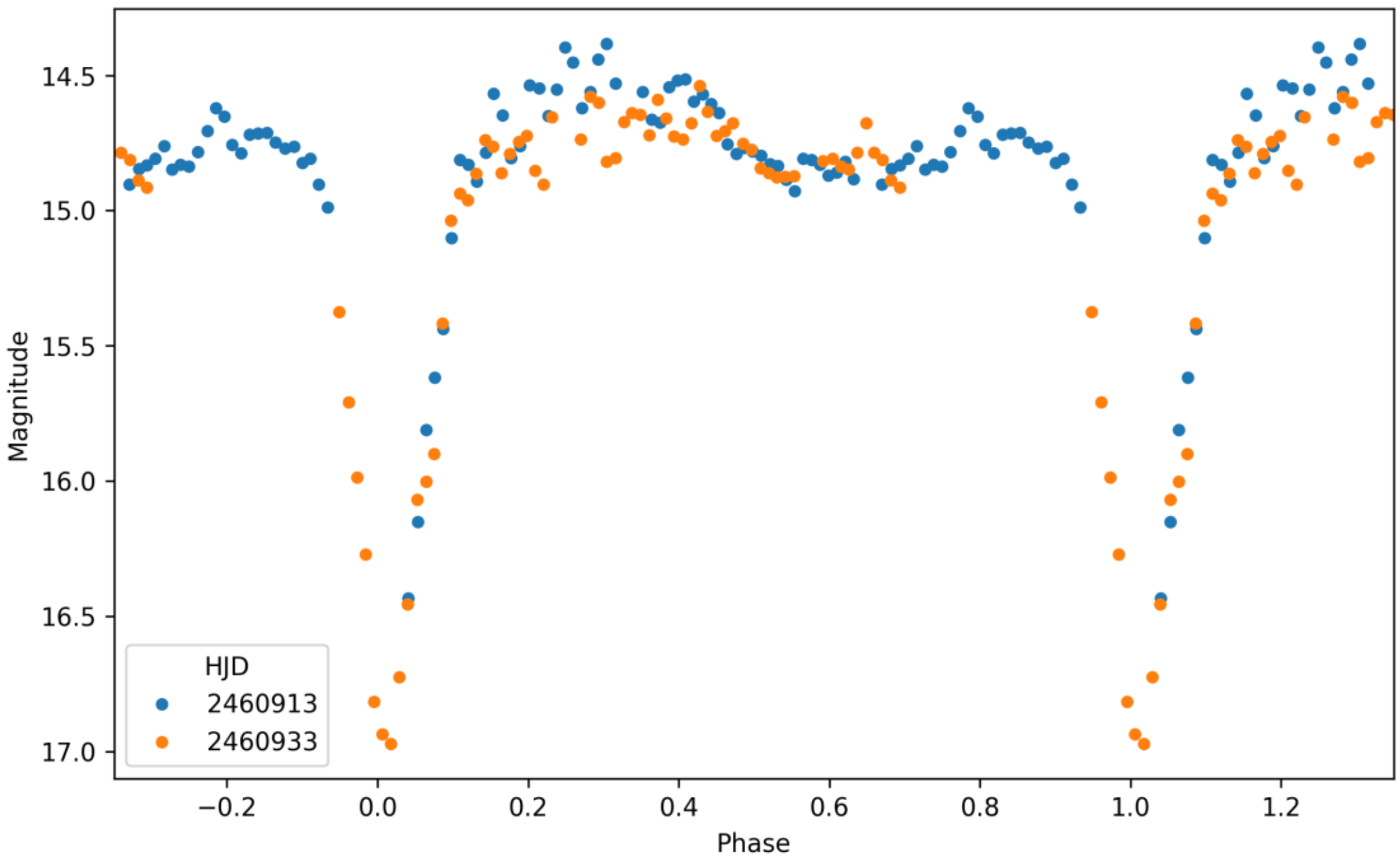

Figure 4: Phased Sloan-r light curves of ZTF J191814.28+010850.2 obtained with Chilescope Telescope 1 on 26 August 2025 and 15 September 2025, folded with the ephemeris given in Sect. 3.

The ZTF and Chilescope observations consistently show a deep eclipse at the derived orbital period. The Chilescope data also provide indications of cycle-to-cycle variations outside eclipse, particularly around phase 0.3. Although listed as an EA-type eclipsing variable in the ZTF catalogue, the very short period, deep eclipse, pronounced out-of-eclipse variability and similar long-term brightness variation in both ZTF bands argue against a straightforward detached-binary interpretation and support an eclipsing cataclysmic-variable candidate interpretation.

## 4 Conclusion

The orbital period, deep eclipse and pronounced short-timescale variability make ZTF J191814.28+010850.2 a promising eclipsing cataclysmic-variable candidate in the SW Sex period range. However, the defining characteristics of SW Sex stars are primarily spectroscopic. Time-resolved spectroscopy is therefore required to establish the nature of the system and to test its possible membership in this class. Further photometric monitoring would also help to characterize the apparent cycle-to-cycle variations and determine whether the long-term brightness variations continue.

## Acknowledgements

This research has made use of the SIMBAD database and the VizieR catalogue access tool, operated at CDS, Strasbourg, France, the International Variable Star Index (VSX) database, operated at AAVSO, Cambridge, Massachusetts, USA, the NASA/IPAC Infrared Science Archive and the SAO/NASA Astrophysics Data System, USA. This work is based in part on observations obtained with the Zwicky Transient Facility. The authors thank the referees for helpful comments.